# Interaction of ammonium with birnessite: Evidence of a chemical and structural transformation in alkaline aqueous medium

Hella Boumaiza[a,b,c], Romain Coustel[b], Christelle Despas[b], Christian Ruby[b], Latifa Bergaoui[a,c,*]

a Laboratoire de Chimie des Matériaux et Catalyse, Faculté des Sciences de Tunis, Université Tunis El Manar, Tunisia
b Laboratoire de Chimie Physique et Microbiologie pour l'Environnement (LCPME)-UMR 7564, CNRS-Université de Lorraine, 405 rue de Vandoeuvre, 54600 Villers-lès-Nancy, France
c Département de Génie Biologique et Chimique, Institut National des Sciences Appliquées et de Technologies (INSAT), Université de Carthage, Tunis, Tunisia

The ammonium cation interaction with Na-birnessite in aqueous alkaline medium was studied. Solution and solid analysis give evidence that birnessite is not only acting as a cationic exchanger toward $NH_4^+$. The surface analysis performed by XPS showed that N1s spectra are characterized by the existence of two different environments: one assignable to an interlayer $NH_4^+$ and the second to a chemisorbed N-species. Structural and chemical transformations were observed on birnessite with nitrogen mass balance deficit. The monitoring of $NH_4^+$, $Na^+$, $Mn^{2+}$, $NO_3^-$ and $NO_2^-$ and solid changes (average oxidation state of Mn, cation exchange capacity, solid nitrogen content and symmetry evolution identified by XRD and FTIR) indicate unambiguously that $NH_4+$ reacts chemically with the birnessite.

## 1. Introduction

Among the various manganese oxides, Birnessite was found to be the most common occurring form in soils and sediments (Bargar et al., 2009). The crystal system of birnessite was shown to be either triclinic or hexagonal [re] but monoclinic system was also sometimes reported [ref]. All the forms consist of a lamellar structure of edge-shared $MnO_6$ octahedra in which the layers are negatively charged and with an interlayer space containing a compensating cation (*e.g.*. $Na^+$, $K^+$, $Mn^{2+}$, $Mn^{3+}$).

In the triclinic form, the charge deficit of the layers is due to the mixed valence states of manganese: +III and +IV (Post and Veblen, 1990; Post et al., 2002). The layers have been described as an alternation of two $Mn^{IV}O_6$ rows and one $Mn^{III}O_6$ along the *b* axis (Drits et al., 1997; Silvester et al., 1997; Lanson, V. A. Drits, et al., 2002). The triclinic symmetry is caused by the presence of the $Mn^{III}O_6$ octahedra distorted by the Jahn Teller effect (Lanson et al., 2000). The manganese cations present in the layers of the hexagonal form are mainly in the oxidation state +IV, the negatively charged layers being mainly caused by the presence of vacancies (Lanson et al., 2000). The charge deficit is compensated by the presence of interlayer manganese in the oxidation state +II and/or +III. These protons are located above and below the vacancies (Silvester et al., 1997). Consequently, different concentrations of $Mn^{II}$, $Mn^{III}$ and $Mn^{IV}$ are present in the two forms of birnessite.

Both forms have been known to participate in a large variety of reactions: ion exchange and sorption reaction (Golden, 1986; Le Goff et al., 1996; Peacock and Sherman, 2007; Jiang et al., 2015), as well as redox reaction with heavy metals, organic and/or (?) inorganic molecules (Scott and Morgan, 1995; Scott and Morgan, 1996; Renuka and Ramamurthy, 2000; Rao et al., 2008; Remucal and Ginder-Vogel, 2014). Given that the birnessite reactivity is influenced by the manganese valence state, determining the $Mn^{II}$, $Mn^{III}$ and $Mn^{IV}$ ratios is of great importance.

It is possible to determine the manganese average oxidation state (AOS) of birnessite by chemical titrations (Lingane and Karplus, 1946; Vetter, K. J. Jaeger, 1966; Murray et al., 1984; Silvester et al., 1997) or by solid characterization via Mn K- and L-edge XANES analysis (Manceau et al., 1992; Chalmin et al., 2009). These methods are bulk techniques that do not provide information about the near-surface chemistry of the oxide where the reactions actually take place.

To this purpose, the X-ray Photoelectron spectroscopy appears to be the technique of choice allowing the determination of the valence state of manganese located near the surface..Despite the existence of many studies that focused on the structural determination of triclinic and hexagonal birnessites, and the conversion from one phase to another (Post and Veblen, 1990; Drits et al., 1997; Silvester et al., 1997; Lanson et al., 2000; Lanson et al., 2002), there are only a few that reported on the determination of $Mn^{II}$, $Mn^{III}$ and $Mn^{IV}$ relative concentrations in birnessite at the surface. The oldest one (Nesbitt and Banerjee, 1998) was based on the fitting of the $Mn2p_{3/2}$ of different K-birnessite. Accordingly, some studies focused on the AOS determination by XPS using the decomposition of the $Mn2p_{3/2}$ line (Yin et al., 2012; McKendry et al., 2015; Boumaiza et al., 2017) but the obtained AOS values measured on the surface were systematically lower then the bulk determination indicating the presence of a $Mn^{III}$ excess on the surface. More recently, Ilton et al., (2016) determined the manganese distribution for several manganese oxide surfaces by comparing the mutiplet splitting of Mn $2p_{3/2}$, 3s and 3p lines. This study revealed that the use of the 3s splitting was not appropriate for birnessite, but the determination of the AOS via the 3p splitting was more accurate in comparison with the charge balancing cations in the interlayer region measured by bulk analysis. In contrast the distribution obtained by the $2p_{3/2}$ splitting which values were systematically more reduced. The authors raised the possibility that the quantification extracted from the analyses of the $Mn2p_{3/2}$ XPS peaks are not precise because of the well-known sensitivity of the peak shape to differences in the Mn bonding environment. They concluded that for an accurate AOS determinations, the Mn(3p) could be combined with the Mn(2p) to provide a greater precision of the Mn speciation in the birnessite.

When submitted to an acidic treatment, the triclinic birnessite undergoes a symmetry conversion to hexagonal (Drits et al., 1997; Silvester et al., 1997; Lanson et al., 2000). This change is the result of the release of $Mn^{II}$ in the solution accompanied by a modification of the $Mn^{II}$, $Mn^{III}$ and $Mn^{IV}$ proportions in the solid. The study of this change by XRD and FTIR analyses is well documented in the literature (Lanson et al., 2000; Ling et al., 2017). But, to the very best of our knowledge, no study focused on the variation of the $Mn^{II}$, $Mn^{III}$ and $Mn^{IV}$ surface proportions by using XPS during this acidic treatment. The main goal of this study is thus to follow the changes of triclinic birnessite placed at pH 3 by monitoring both the solid and liquid (properties. The reliability to use XPS for the determination of the Mn Average Oxidation State (AOS) the near-surface will be evaluated and compared to bulk analysis.

## 2. Materials and methods

### 2.1. Chemicals

All chemicals were purchased from Sigma-Aldrich. Manganese(II) chloride tetrahydrate ($MnCl_2,4H_2O$, ACS reagent, $\geq 98\%$), sodium permanganate monohydrate ($NaMnO_4,H_2O$, ACS reagent $\geq 97\%$) and sodium hydroxide (NaOH, BioXtra, $\geq 98\%$ pellets anhydrous) were used for birnessite's synthesis. A hydrochloridric acid solution (0.2 mol $L^{-1}$) was used to stabilize the birnessite suspension at the desired pH.

### 2.2. Synthesis of birnessite

The triclinic birnessite was obtained following the alkaline method of Boumaiza et al., (2017) in which 125 mL of NaOH (8.8 mol $L^{-1}$) is added dropwise during 2 hours to a mixture of 250 mL of $NaMnO_4$ (0.1 mol $L^{-1}$) and 125 mL of $MnCl_2$ (0.6 mol $L^{-1}$). The mixture is then mixed during ? an additional 30 minutes, aged at 60 °C during 14h, washed until the supernatant pH is 9 – 10 and dried at 60 °C during 16 hours. The obtained product consisted on triclinic Na-birnessite with a general formula $[Mn^{III}_{0.35} Mn^{IV}_{0.65} O_2][Na^+_{0.35} 0.7\ H_2O]$, an Average Mn Oxidation State (AOS) equal to 3.65 and a Cation Exchange Capacity (CEC) equal to (2.3 $\pm$ 0.1) meq $g^{-1}$.

The hexagonal birnessite was obtained from the transformation of Na-birnessite in acid solution: Na-birnessite suspension (6 $gL^{-1}$) was stabilized at pH 3 by adding hydrochloridric acid solution (0.2 mol $L^{-1}$). An additional experiment without pH stabilization was also performed (pH = 10). For the liquid analysis, samples were collected at different contact times and filtered through 0.22 µm syringe filter. After 24 h of contact time, the remaining birnessite was collected for solid state characterizations.

### 2.3. Characterization techniques

Powder X-ray Diffraction (XRD) was performed at room temperature by placing the powders on zero background quartz sample holders and by using an X'Pert MPD diffractometer (Panalytical AXS) and an X'Celerator as a detector. The goniometer radius is 240 nm and the divergence slit module is fixed (1/2° divergence slit, 0.04 rd Sollers slits). The XRD patterns are recorded using $CuK_\alpha$ radiation ($\lambda$=0.15418 nm).

FTIR analysis was performed using a Bruker Vertex 70 v equipped with a DLaTGS detector. The transmission mode was used to record spectra in which KBr pellets contained 2 – 5 mg of sample. In the 5000 – 200 $cm^{-1}$ wavenumber range, 100 scans were collected for each sample with 4 $cm^{-1}$ resolution. The background spectrum of KBr was recorded in the exact same conditions.

X-ray photoelectron spectra were recorded on a KRATOS Axis Ultra X-ray photoelectron spectrometer (Kratos Analytical, Manchester, UK) equipped with a monochromated Al K$\alpha$ source (h$\upsilon$ =1486.6 eV, spot size 0.7 mm×0.3 mm). The samples were pressed onto a Cu tape fixed on a holder and introduced into the spectrometer. The detector is a hemispherical analyzer at an electron emission angle of 90° and pass energy of 160 eV (survey spectra) or 20 eV (core level spectra). For the core-level spectra, the overall energy resolution, resulting from monochromator and electron analyzer bandwidths, was 800 meV. Charge correction was carried out using the C 1s core line, setting adventitious carbon signal (H/C signal) to 284.6 eV.

Mn content in the supernatant was determined using Inductive Coupled Optical Emission Spectroscopy ICP-AES (Jobin Yvon-Horiba, Ultima).

The Mn AOS of the bulk birnessite was measured by potentiometric titration using a Mohr salt, potassium permanganate and sodium pyrophosphate dibasic (Lingane and Karplus, 1946; Vetter, K. J. Jaeger, 1966).

## 3. Results and discussion.
### 3.1. Analysis of soluble manganese

It is first necessary to point out that, the manganese in solution is assumed to be Mn(II) since Mn(III) and Mn(IV) are esssentially stable in solid-phase (Langston et al., 2010).

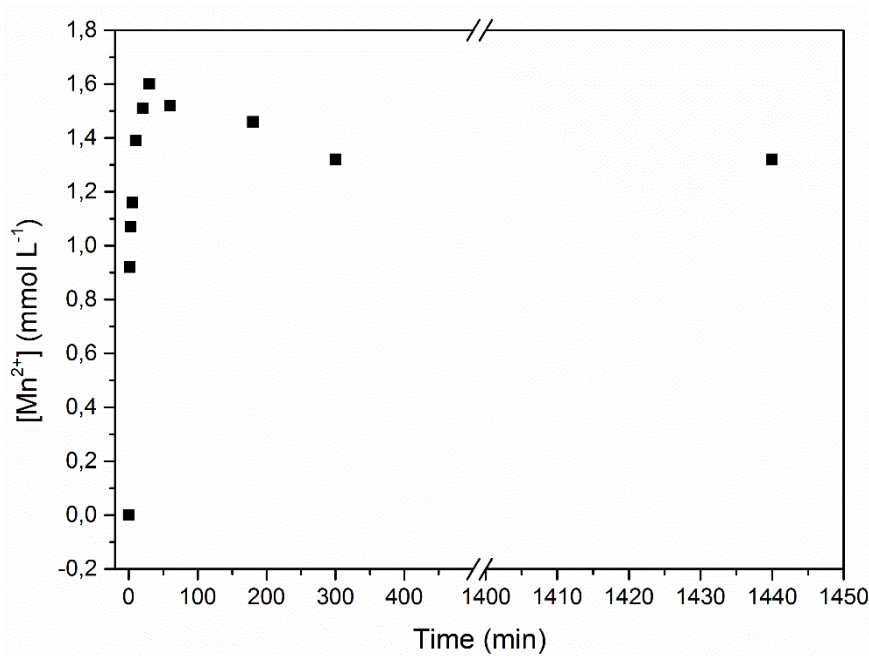

**Fig. 1**. Evolution of Mn(II) concentration in function of time at pH 3 ([Na-bir] = 6 g L$^{-1}$, T = 25 °C, reaction time = 24 h)

The Mn(II) content in the supernatant at pH 3 as a function of time is shown in Fig. 1. The trend/ evolution (?) is characterized by two distinct steps: during the first 30 minutes, the Mn(II) concentration rises rapidly to a maximum of 1.6 mmol L$^{-1}$. Then, the concentration decreases gradually to 1.32 mmol L$^{-1}$ at 180 minutes and this value is retained until 24 h reaction time. At this pH, the amount of Mn(II) released in the solution only represents 3% of Na-birnessite initially introduced in the solution while at pH 10, no Mn(II) was detected in the supernatant. This is consistent with the previous observation showing thatthe reaction of r triclinic birnessite placed in an acidic medium (Silvester et al., 1997) could be explained by a two steps reaction. The first step corresponds to the disproportionation of Mn(III) in the layers into Mn(II) and Mn(IV), the Mn(II) formed being then dissolved in acidic medium by creating vacancies in the layers by following reaction (1):

$$2 \; Mn^{3+}{}_{layer} \rightarrow Mn^{4+}{}_{layer} + Mn^{2+}{}_{layer} \rightarrow Mn^{4+}{}_{layer} + vacancy + Mn^{2+}{}_{aq} \quad (1)$$

In a second step, a part of the dissolved Mn(II) re-adsorbs on the surface, above and below vacancies, resulting in the diminution of the amount of aqueous Mn(II) (Silvester et al., 1997). Another study also suggested that the absorbed Mn(II) is able to react with Mn(IV) ions in the layers to regenerate Mn(III) (Zhao et al., 2016)

### 3.2. Solid analysis
- *<u>XRD and FTIR characterizations</u>*

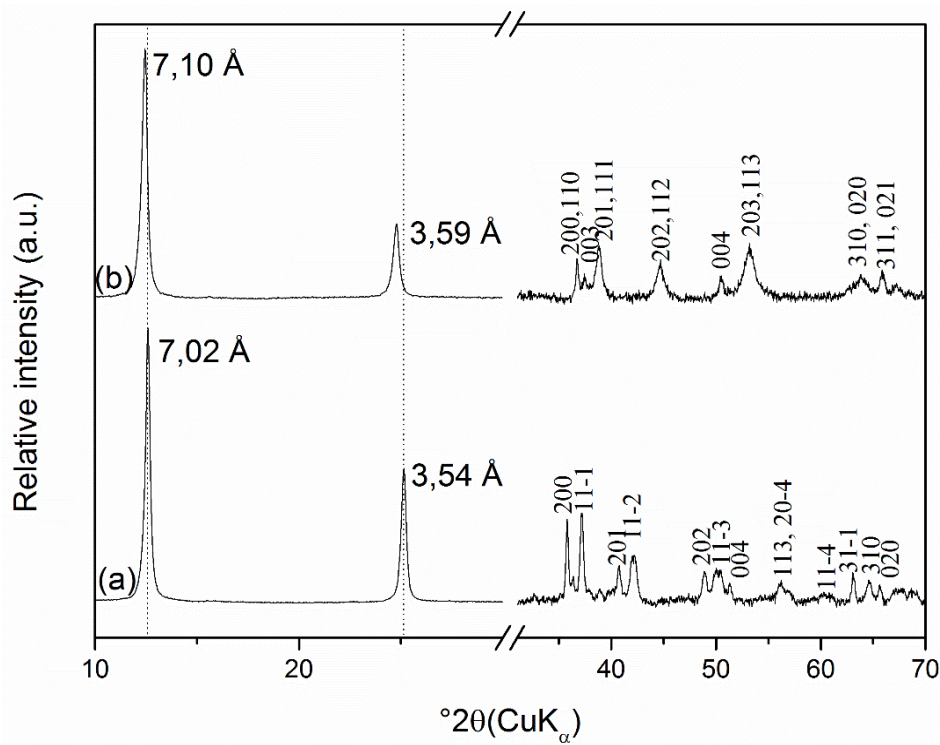

**Fig. 2**. XRD patterns of a) triclinic Na-birnessite and b) hexagonal H-birnessite ([Na-bir] = 6 g L$^{-1}$, T = 25 °C, reaction time = 24 h, pH= ?)

The XRD patterns of the triclinic Na-birnessite is presented in figure 2a in comparison to the one maintained at pH 3 during 24 h (Fig. 2b).

For 2θ range from 10 to 30 °, the two XRD patterns are quite similar and are both characterized by the presence of two main peaks at around 7 and 3.5 Å assigned respectively to the basal reflections 001 and 002. In the triclinic form, the two main peaks are located at angles corresponding to the distances 7.02 and 3.54 Å. When submitted to acidic treatment, a slight shift toward lower 2θ values is observed and the two main peaks correspond henceforth to the distances 7.10 and 3.59 Å. For 2θ range from 30 to 70 °, the initial reflections of the triclinic form collapse when submitted to acidic treatment and the product maintained at pH 3 during 24 h presents other reflection peaks corresponding to hexagonal birnessite (Drits et al., 1997; Lanson et al., 2000).

The solids were further characterized by FTIR analysis and the spectra are presented in figure 3. The initial triclinic Na-birnessite is characterized by the presence of bands located at 245, 360, 418, 478, 512 and 634 cm$^{-1}$ with a shoulder at 550 cm$^{-1}$. These bands are assigned to the Mn-O lattice vibration, in good agreement with the reported spectra of Na-birnessite (Potter and Rossman, 1979; Feng, 2005; Cui et al., 2009; Ling et al., 2017).

The spectrum of the birnessite maintained at pH 3 during 24 h is sensibly/clairly different and is only characterized by the presence of two large bands located at 430 and 495 cm⁻¹ characteristics of the hexagonal symmetry (Johnson and Post, 2006; Ling et al., 2017).

The liquid and solid analyses thus showed that, a triclinic Na-birnessite is converted into hexagonal H-birnessite when maintained at pH 3 during 24 h, as it has been previously observed (Drits et al., 1997; Silvester et al., 1997; Lanson et al., 2000).

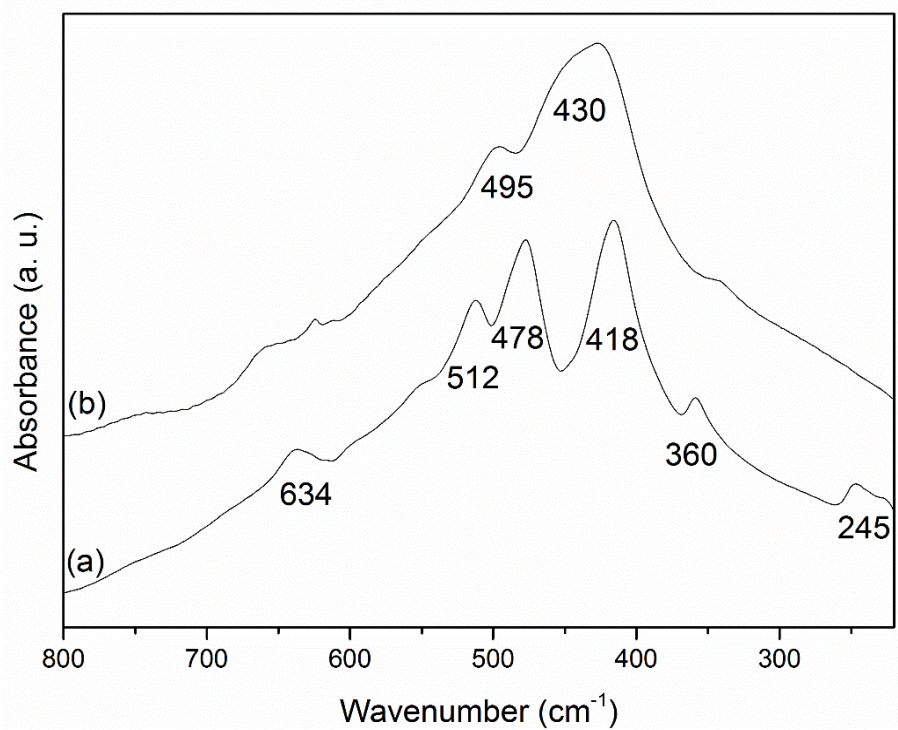

**Fig. 3**. FTIR spectra of a) triclinic Na-birnessite and b) hexagonal H-birnessite ([Na-bir] = 6 g L⁻¹, T = 25 °C, reaction time = 24 h, pH = ?).

- *XPS analysis*

Surface properties of Na-birnessite and H-birnessite were examined by XPS. Overview XPS spectra (see supplementary materials) show core-level photoelectron peaks at ~50 eV (Mn3p), ~285 eV (C1s), ~530 eV (O1s), ~642 eV (Mn2p$_{3/2}$), ~654 eV (Mn2p$_{1/2}$), ~770 eV (Mn2s) and ~1071 eV (Na1s). The C1s peak should be attributed to a hydrocarbon contamination of the birnessite surface before introduction into the XPS chamber.

The latter peak disappears for the hexagonal form, indicating the total exchange of the interlayer sodium.

Peculiar attention was paid to Mn2p 3/2 and Mn3p peaks from which it has been claimed that the valence state of Mn can be extracted [citer Nessbitt et Ilton]. The Mn2p3/2 spectrum of Na-birnessite is centered at 642 eV with shoulders at 641 eV and 643 eV, as expected for Mn(III)-Mn(IV) birnessite [citer Boumaiza et al 2017]. Similar, but widened, feature is observed for H-birnessite (Figure 4.a) with additional shoulder at 639 eV. This low binding energy component is attributed to Mn(II) contribution [citer Nessbitt et Ilton]. Then XPS results confirm that H-birnessite presents Mn(II) according to the mechanism of Na-birnessite conversion in acidic medium proposed by Silvester et al. [citer la ref].

Nesbitt et al. have proposed a method to isolate the Mn(II), Mn(III) and Mn(IV) contributions of Mn2p 3/2 peak observed for binessite : based on the earlier work of Gupta and Sen [Gupta, RP and Sen, S.K. (1974) According to calculations of multiplet structure core p-vacancy levels Physical Reviews B, 10,7119] each valence state contribution is modeled as a multiplet whose parameters are derived from ab-initio calculations. Mn2p 3/2 spectra of Na-birnessite and H-birnessite were fitted following the recommendations of Nesbitt et al. and corresponding results are given in Figure 4 and Table 1. This analysis confirms that Na-birnessite presents only Mn(III) and Mn(IV) while H-birnessite presents Mn(II). It appears also that Mn(III) contain decreases while Mn(IV) contain increases after acidic treatment, as expected from the mechanism proposed by Silvester et al.[ref].

Recently, Ilton et al., (2016) claimed that a more reliable determination of Mn(II), Mn(III) and Mn(IV) distribution on the surface can be obtained from the Mn3p signal. Corresponding experimental results are showed in Figure 4: Mn3p peaks for Na-birnessite and H-birnessite are located à 50 eV with a shoulder at 48 eV. These authors proposed an empirical approach where each valence state contribution is described as a multiplet whose shape is experimentally determined from XPS measurement on monovalent Mn oxide (MnO, MnOOH and MnO$_2$). Results obtained from such Mn3p fitting procedure are given in Figure 4 and table 1. As determined from the Mn2p 3/2 spectra, the Ilton et al's procedure indicate that Mn(II) appears after acid treatment while Mn(III) and Mn(IV) contain decreases and increases, respectively.

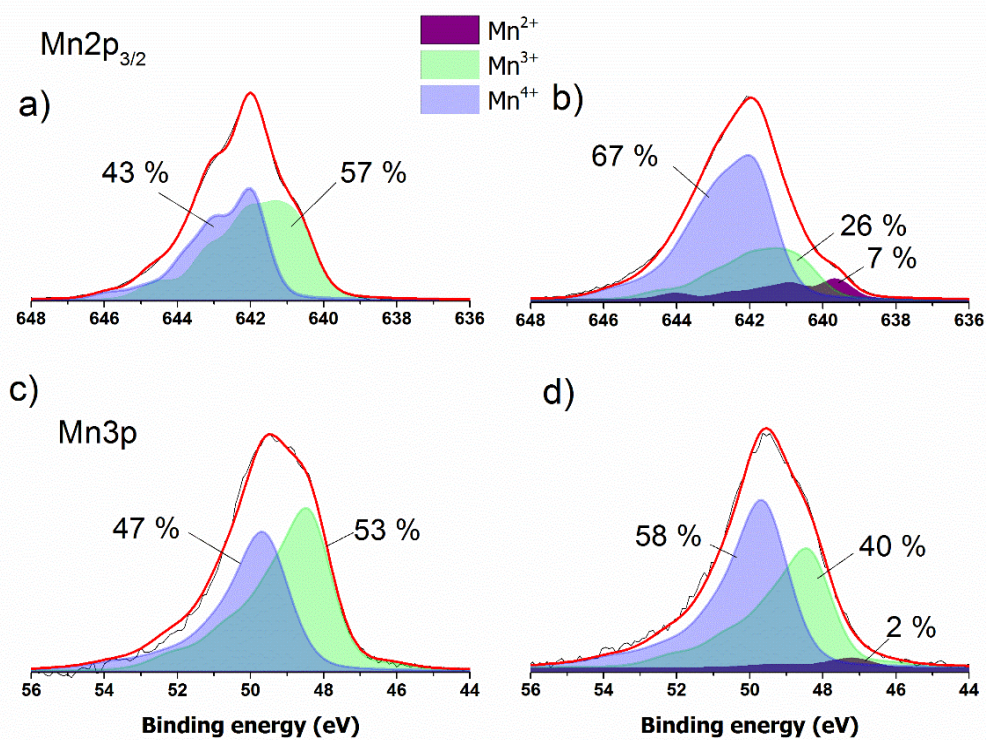

**Fig.4**. Mn2p$_{3/2}$ and Mn3p XPS spectra of the triclinic Na-birnessite (a & c) and the hexagonal H-birnessite (b & d) ([Na-bir] = 6 g L$^{-1}$, T = 25 °C, reaction time = 24 h, pH= 3).

The quantitative analysis shows that the Na/Mn ratio decreases from 0.5 for the triclinic form to 0 for the hexagonal one indicating that all the initial sodium was released in the solution after acidic treatment.

The determination of Mn$^{II}$, Mn$^{III}$ and Mn$^{IV}$ distribution on the surface based on the Mn2p$_{3/2}$ and Mn3p are summarized in Table 1 in comparison to the bulk determination of the Mn AOS. The Mn AOS determined by XPS analysis of both forms of birnessite are similar for the two fitting methods. Indeed, the triclinic form presented aMn AOS of 3.43 and 3.47 for the Mn2p$_{3/2}$ and Mn3p fittings respectively. The same applies to the hexagonal form which Mn AOS at the surface were equal to 3.60 and 3.56 for the Mn2p$_{3/2}$ and Mn3p fittings respectively. These obtained values from surface analysis are considerably more reduced than the Mn AOS determined by bulk analysis corresponding to 3.65 and 4.00 for the triclinic and hexagonal forms respectively.

**Table 1**: $Mn^{II}$, $Mn^{III}$ and $Mn^{IV}$ distribution on the surface compared to the bulk determination

|  | **Triclinic Na-birnessite** | | **Hexagonal H-birnessite** | |
|---|---|---|---|---|
|  | $Mn2p_{3/2}$ fitting | Mn3p fitting | $Mn2p_{3/2}$ fitting | Mn3p fitting |
| **Mn(II) (%)** | 0 | 0 | 7 | 2 |
| **Mn(III) (%)** | 57 | 53 | 26 | 40 |
| **Mn(IV)** | 43 | 47 | 67 | 58 |
| **AOS$_{surface}$** | 3.43 | 3.47 | 3.60 | 3.56 |
| **AOS$_{bulk}$** | 3.65 | | 4.00 | |

However, it is necessary to point out that the triclinic form presented systematically a Mn AOS more reduced than the hexagonal form, in both bulk and surface analyses, in good agreement with the proposed conversion mechanism between the two forms (Lanson et al., 2000). Indeed the symmetry conversion is the result of the exchange of the interlayer $Na^+$ by $H_3O^+$ from the solution and the disproportionation of Mn(III) to Mn(IV) in the layers and Mn(II) in the solution. The hexagonal birnessite thus presents higher amounts of Mn(IV) and consequently a higher Mn AOS value.

The use of the $Mn2p_{3/2}$ fitting combined with the Mn3p gave systematically lower Mn AOS values than the bulk, unlike what has been observed in the study of Ilton et al. (2016). This suggests that manganese at the near-surface of all types of birnessite is significantly more reduced than in the bulk.

## 4. Conclusion

The conversion of the triclinic birnessite into hexagonal one was monitored using liquid and solid characterization. The ICP, FTIR and XRD characterization were similar to what have been previously observed. The determination of $Mn^{II}$, $Mn^{III}$ and $Mn^{IV}$ distribution on the surface by XPS based on the $Mn2p_{3/2}$ and Mn3p fitting parameters gave similar results that were systematically more reduced than the bulk analysis. The XPS analyses appears to be not an accurate tool to monitor and determine the bulk Mn AOS in the birnessites.